\documentclass[%
 reprint,
superscriptaddress,
 amsmath,amssymb,
 aps,
 pre,
floatfix,
]{revtex4-2}
\usepackage{cancel}
\usepackage[caption=false]{subfig}
\usepackage{graphicx}
\usepackage{dcolumn}
\usepackage{bm}
\usepackage[hidelinks]{hyperref}
\usepackage{xcolor}
\usepackage[normalem]{ulem}
\definecolor{ra}{rgb}{0.8, 0.0, 0.0}

\begin{document}

\preprint{APS/123-QED}

\title{Reducing finite-size effects with reweighted renormalization group transformations}
\author{Dimitrios Bachtis}
\email{dimitrios.bachtis@phys.ens.fr}

\affiliation{Laboratoire de Physique de l'Ecole Normale Sup\'erieure, ENS, Universit\'e PSL,
CNRS, Sorbonne Universit\'e, Universit\'e de Paris, F-75005 Paris, France}

\include{ms.bib}

\date{June 05, 2023}

\begin{abstract}
We combine histogram reweighting techniques with the two-lattice matching Monte Carlo renormalization group method to conduct computationally efficient calculations of critical exponents on systems with moderately small lattice sizes. The approach, which relies on the construction of renormalization group mappings between two systems of identical lattice size to partially eliminate finite-size effects, and the use of histogram reweighting to obtain computationally efficient results in extended regions of parameter space, is utilized to explicitly determine the renormalized coupling parameters of the two-dimensional $\phi^{4}$ scalar field theory and to extract multiple critical exponents. We conclude by quantifying the computational benefits of the approach and discuss how reweighting opens up the opportunity to extend Monte Carlo renormalization group methods to systems with complex-valued actions.
\end{abstract}

\maketitle

\section{\label{sec:level1}Introduction}

The renormalization group~\cite{PhysicsPhysiqueFizika.2.263,PhysRevB.4.3174,PhysRevLett.28.240,WILSON197475,RevModPhys.47.773} provides a quantitative framework to investigate phase transitions and has therefore significantly impacted research pertinent to these ubiquitous phenomena which arise in distinct research fields. Traditional applications of real-space renormalization group methods are highly successful in producing accurate results when compared against theory or experiment, but they are simultaneously hindered by the systematic errors introduced through the application of a real-space transformation which truncates the original degrees of freedom according to a predefined rule. 

In contrast, Monte Carlo simulations of statistical systems, which are introduced based on the theory of Markov processes, are able to guarantee a certain form of exactness. Within the Monte Carlo approach, it is statistical errors that emerge, and which can be dealt with accordingly. Nevertheless, Monte Carlo methods are hindered by the critical slowing down effect~\cite{newmanb99}, a problem which arises as one simulates systems of increasing lattice size in the vicinity of the critical point. One anticipates that via the combination of real-space renormalization group methods and Monte Carlo simulations, it might be possible to simultaneously mitigate both the systematic errors introduced by the renormalization group transformations, as well as the critical slowing down effect which emerges within the Monte Carlo approach. It is exactly this perspective that motivates implementations of the Monte Carlo renormalization group~\cite{PhysRevLett.37.461,PhysRevLett.42.859}.

\begin{figure}
\center
\includegraphics[width=8.2cm]{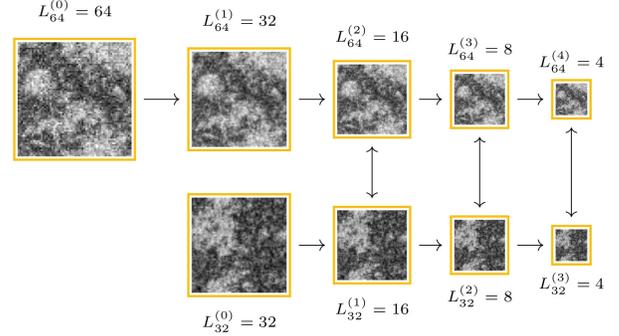}
\caption{\label{fig:illust} Illustration of the two-lattice matching Monte Carlo renormalization group method where calculations are conducted on two systems of identical lattice size. We remark that it is additionally possible to conduct renormalization group calculations between an original and a renormalized system, namely $L_{32}^{(0)}$ and $L_{64}^{(1)}$. }
\end{figure}

Monte Carlo renormalization group methods were introduced by Ma~\cite{PhysRevLett.37.461}. Significant contributions towards increasing their usability and efficiency were proposed by Swendsen~\cite{PhysRevLett.42.859}. In this method one relies on the construction of the linearized renormalization group transformation matrix to introduce additional terms within the calculation, thus enabling a certain control over the emergent errors. In constrast to straightforward Monte Carlo approaches which are generally implemented using  a finite-size scaling analysis based on systems of increasing lattice size, the accuracy obtained within the context of the Monte Carlo renormalization group can be systematically improved via the application of iterative transformations which instead reduce the lattice size of the system. These transformations drive the system to the fixed point if it originally resides sufficiently close to the phase transition~\cite{Burkhardt1982}.

We remark that computational implementations of the renormalization group are not severely affected by finite-size effects in comparison to other methods, such as finite-size scaling extrapolations. This is due to the fact that calculations of critical exponents in the renormalization group are conducted on exclusively two lattice sizes.  In fact, finite-size effects can be  reduced further in the renormalization group by conducting the calculations on two renormalized systems of identical lattice size, an approach often named as the two-lattice matching Monte Carlo renormalization group~\citep{swecargese,Wilsonbook,PhysRevB.29.4030,BOWLER1986375,PhysRevLett.108.061601}.

 Nevertheless, in the aforementioned method, the calculation of certain critical exponents necessitates the accurate determination of the renormalized coupling parameters. Without prior insights of the critical point, the accurate determination of the renormalized coupling parameters is achieved by constructing an equivalence between observables of two renormalized systems whose correlation length differs by a predefined rescaling factor. This two-lattice matching approach requires a large number of direct Monte Carlo simulations for two distinct systems in extended regions of parameter space and it becomes a computationally expensive task.

In this manuscript, we combine real-space renormalization group methods with histogram reweighting~\cite{PhysRevLett.61.2635} to determine, with computational efficiency and by partially eliminating finite-size effects, the renormalized coupling parameters of the two-dimensional $\phi^{4}$ scalar field theory. In addition, we use histogram reweighting and the renormalization group to study the explicit symmetry-breaking of the $\phi^{4}$ theory without the need to simulate actions which include explicit symmetry-breaking terms. We therefore explore if we can provide reweighting-enhanced calculations of the exponents associated with the relevant variables of the renormalization group. We conclude by quantifying the computational benefits provided by the incorporation of histogram reweighting in the two-lattice matching Monte Carlo renormalization group method and discuss how reweighting opens up the opportunity to extend Monte Carlo renormalization group methods to systems with complex-valued actions.

\section{Renormalization group and the $\phi^{4}$ theory}

In this manuscript, we are interested in applying renormalization group transformations to conduct calculations of critical exponents based on two systems of identical lattice size. We therefore briefly review the fundamental concepts pertinent to numerical implementations of the renormalization group. We begin by considering an original system that undergoes a second-order phase transition and which is described by a lattice size $L_{0}$ in each dimension. 

The application of renormalization group transformations on the original system with $L_{0}$ will produce a renormalized system with lattice size $L'$, given by:
\begin{equation}
L'=\frac{L_{0}}{b^{n}}.
\end{equation}
where $L'$ is the renormalized lattice size after $n$ consecutive transformations and $b$ is the rescaling factor.  In this manuscript we consider that $b=2$, hence halving the lattice size in each dimension with the application of a renormalization group transformation.  To avoid ambiguity when referring to renormalized systems obtained from different original lattice sizes $L_{0}$ after $n$ iterative renormalization group transformations we will use the notation $L^{(n)}_{L_{0}}$. 

If we assume that the original system resides in the vicinity of a second-order phase transition then the reduction of the original lattice size $L_{0}$ implies an analogous reduction of the original correlation length $\xi_{0}$. The renormalized correlation length $\xi'$ is then equal to:
\begin{equation}
\xi'=\frac{\xi_{0}}{b^{n}}.
\end{equation}

We recall that the correlation length is a quantity which arises dynamically as one approaches the critical point. Since the original and the renormalized systems are described by different correlation lengths $\xi_{0}$ and $\xi'$ then this implies that the two systems have a different distance from the critical point.

The distance from the critical point $K_{c}$ can be quantified via the reduced coupling constants $t_{0}$ and $t'$:
\begin{equation}
t_{0}=\frac{K_{c}-K_{0}}{K_{c}}, 
\end{equation}
\begin{equation}
 t'=\frac{K_{c}-K'}{K_{c}},
\end{equation}

 At the critical value of the fixed point $K_{c}$ the correlation lengths $\xi_{0}$ and $\xi'$ of the original and the renormalized systems diverge and intensive observable quantities $O_{0}$ and $O'$ become equal:
\begin{equation}\label{eq:n1}
O_{0}(K_{c})=O'(K_{c}).
\end{equation}

The intersection point of the two observables serves as an estimate of the critical fixed point $K_{c}$ of the system. 

 We remark that, while the observation of Eq~\ref{eq:n1} is mathematically valid , it is subject to the presence of systematic errors which emerge by finite-size effects, the iterations of the renormalization group transformation, and the initial distance of the system from the critical point. This implies that, in numerical implementations, intensive observables might not always intersect. To mitigate this problem we will conduct calculations based on two renormalized systems of identical lattice size to reduce finite-size effects.
 
In this manuscript, we consider the two-dimensional $\phi^{4}$ scalar field theory, which is discretized on a square lattice, and is described by the Euclidean action~\cite{PhysRevD.103.074510}:
\begin{equation}\label{eq:midaction} \nonumber
S = -\kappa \Big(\sum_{\langle ij \rangle}\phi_{i} \phi_{j}-2\sum_{i} \phi_{i}^{2}\Big) + \frac{\mu^{2}}{2} \sum_{i} \phi_{i}^{2} +  \frac{\lambda}{4}  \sum_{i}\phi_{i}^{4},
\end{equation}
where $\kappa,\mu^{2},\lambda$ are dimensionless parameters and $\langle ij \rangle$ denotes nearest-neighbors. For the case of the $\phi^{4}$ theory described above, the second-order phase transition is induced by varying the squared mass $\mu^{2}$, hence $K_{0}\equiv\mu^{2}$, $K'\equiv\mu'^{2}$ and $K_{c}\equiv \mu_{c}^{2}$. Specifically, for fixed $\lambda>0$, $\kappa>0$ the system undergoes a phase transition between a symmetric and a broken-symmetry phase for a unique value of $\mu_{c}^{2}<0$~\cite{Milchev1986}. The action $S$ can be factorized as 
\begin{equation}
S= \sum_{k=0}^{2} g_{k} S^{(k)},
\end{equation}
where $g_{k}$ corresponds to each coupling constant $\kappa,\mu^{2},\lambda$ and $S^{(k)}$ is the corresponding action term. We have expressed the action in this form to simplify histogram reweighting calculations that will be discussed subsequently.

To apply the renormalization group on the $\phi^{4}$ theory we implement a linear transformation~\citep{PhysRevD.35.672} which produces a rescaled degree of freedom $\phi'$ as:
\begin{equation}
\phi'=\frac{f}{2^{d}} \sum_{i \in \rm{block}} \phi_{i},
\end{equation}
where $d=2$ is the dimension of the system and the factor $f$ is optimized~\citep{PhysRevD.35.672,PhysRevLett.52.2321} in order to produce a critical fixed point. 

The optimization of the transformation is achieved by varying $f$ until an intersection of observables between a renormalized and an original system is obtained. The procedure indicates that an approximately nearest-neighbor renormalized action can be obtained for the discussed system. This optimization process, conducted by searching for an intersection between the original and the renormalized magnetization, yields the value $f=1.09$. We remark that linear transformations are limited in their expressivity and systematic errors emergent from the implementation of a linear transformation are not considered in this manuscript. For an alternative nonlinear transformation of the $\phi^{4}$ theory that does not require the determination of a factor $f$, see Ref.~\cite{PhysRevLett.128.081603}. 

We remark that the consideration of only one operator in the optimization process can result in the conception of an imperfectly optimized transformation. Potential research directions to further improve the quality of the transformation would consider the simultaneous optimization of multiple operators to observe if the emergent critical fixed point is consistent. Alternative directions to improve the transformation would consider the investigation of systematic uncertainties pertinent to the imperfect tuning of $f$ since slight perturbations of $f=1.09$ can produce comparable results. 
 
The intensive observables to be considered in this manuscript are the absolute value of the magnetization:
\begin{equation}
m \equiv |m|= \frac{1}{V} \Big| \sum_{i}\phi_{i} \Big|,
\end{equation}
and the nearest-neighbor interaction:
\begin{equation}
O_{\phi_{i} \phi_{j} }=  - \frac{1}{V} \sum_{\langle ij \rangle} \phi_{i} \phi_{j}.
\end{equation}
where $V=L \times L$ is the size of the system and $\langle ij \rangle$ denotes nearest-neighbors. The minimally correlated configurations of the $\phi^{4}$ theory used for calculations are sampled with a combination of the Metropolis and the Wolff algorithms~\cite{PhysRevLett.62.1087,PhysRevD.79.056008,PhysRevD.58.076003, PhysRevLett.62.361}. The statistical errors are calculated with a binning error analysis technique. 

We briefly describe the implementation of the two-lattice matching Monte Carlo renormalization group, see  Fig.~\ref{fig:illust}. We begin by simulating two $\phi^{4}$ scalar field theories for lattice sizes $L=64$ and $L=32$ and $\mu_{0}^{2}=-0.9515$, $\kappa=1$, $\lambda=0.7$, which is a set of coupling constants that defines a system which resides in the vicinity of the phase transition~\cite{PhysRevD.79.056008}.  We then apply a renormalization group transformation on the system with lattice size $L=64$ to obtain a renormalized system of $L_{64}^{(1)}=32$. It is now possible to implement the two-lattice matching Monte Carlo renormalization group method by comparing observables between the renormalized $L_{64}^{(1)}=32$ and the original $L=32$ system. However to approach the renormalized trajectory the previous step is omitted. Instead we continue by applying consecutive renormalization group transformations on the configurations of $L_{64}^{(1)}=32$ and $L=32$ until we obtain lattices of size $L_{64}^{(4)}=4$ and $L_{32}^{(3)}=4$.  It is now possible to proceed to locate the critical fixed point and to extract the critical exponents. However, we will first explore how reweighting can enhance and simplify calculations pertinent to the two-lattice matching Monte Carlo renormalization group method.

\section{Reweighting of renormalized observables}

Our aim is to implement reweighting to construct mappings between observables of two renormalized systems of identical lattice size. These mappings correspond to a function which relates the renormalized coupling parameters, specifically the renormalized squared masses, in extended regions of parameter space. To introduce the reweighting approach we start by defining the expectation value $\langle O \rangle$ of an arbitrary observable for the original system with lattice size $L_{0}$ as calculated in a Monte Carlo simulation:
\begin{equation} \label{estimo}
\langle O \rangle=\frac{\sum_{i=1}^{N} O_{\sigma_{i}} \tilde{p}_{\sigma_i}^{-1}  \exp[-S_{\sigma_{i}} ]}{\sum_{i=1}^{N} \tilde{p}_{\sigma_i}^{-1}  \exp[ -S_{\sigma_{i}}]},
\end{equation}
where $\tilde{p}_{\sigma_{i}}$ are the sampling probability distributions of a configuration $\sigma_{i}$ and $N$ is the number of Monte Carlo samples. We now select $\tilde{p}_{\sigma_{i}}$ as the probabilities which correspond to the action $S$ of the original system with coupling constants $g_{0} = \kappa$, $g^{(0)}_{1} = \mu_{0}^{2}$, $g_{2} = \lambda$ and lattice size $L_{0}$, and substitute to the above expression to obtain:
\begin{equation} \label{estim}
\langle O \rangle=\frac{\sum_{i=1}^{N} O_{\sigma_{i}}  \exp[-\frac{1}{2}(\mu^{2}-\mu_{0}^{2})S_{\sigma_{i}}^{(1)} ]}{\sum_{i=1}^{N}  \exp[-\frac{1}{2}(\mu^{2}-\mu_{0}^{2})S_{\sigma_{i}}^{(1)} ]}.
\end{equation}

We emphasize, that since we are interested in extrapolating exclusively along the direction of the squared mass $g_{1} = \mu^{2}$ to induce the phase transition, the remaining coupling constants $g_{0} = \kappa$ and $g_{2} = \lambda$ are fixed and have exactly the same value. Consequently, the corresponding action terms $S^{(0)}$ and $S^{(2)}$ cancel in the above expression. So far, we have introduced Eq.~(\ref{estim}) which enables the extrapolation of an observable $O$ of the original system to different values of the squared mass $\mu^{2}$: this is the conventional histogram reweighting setting. 

We are now interested in a different use of histogram reweighting, namely in extrapolating a renormalized observable $O'$ to different values of the squared mass $\mu^{2}$ based on the original probability distribution of the original system. To clarify, we observe that, due to the application of the renormalization group transformation, there exists a mapping between each original configuration $\sigma_{i}$ and the renormalized configuration $\sigma'_{i}$. Consequently, observables of the renormalized system remain, probabilistically, as observables of the original system and can be extrapolated in parameter space using the original probability distribution. Explicitly, the reweighting expression for a renormalized observable is:
\begin{equation} \label{estimm}
\langle O' \rangle=\frac{\sum_{i=1}^{N} O'_{\sigma'_{i}}  \exp[- \frac{1}{2}(\mu^{2}-\mu_{0}^{2})S_{\sigma_{i}}^{(1)} ]}{\sum_{i=1}^{N}  \exp[- \frac{1}{2}(\mu^{2}-\mu_{0}^{2})S_{\sigma_{i}}^{(1)} ]}.
\end{equation}

\begin{figure}
\center
\includegraphics[width=8.cm]{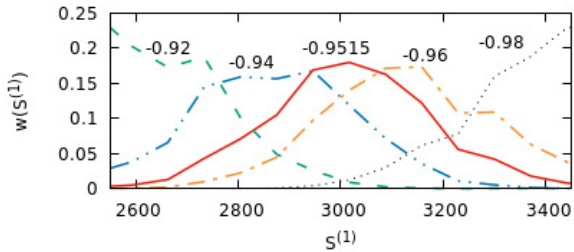}
\caption{\label{fig:fighist} Weight factors $w(S^{(1)})$ versus action $S^{(1)}$ for values of the squared mass $\mu^{2}=-0.98,-0.96,-0.9515,-0.94,-0.92$. We remark that the original simulation is conducted for squared mass $\mu^{2}_{0}=-0.9515$.}
\end{figure}

The important observation is that for each original configuration $\sigma_{i}$ we substitute the action term $S^{(1)}_{\sigma_{i}}$ as calculated on $\sigma_{i}$ but substitute $O'$ as calculated on the corresponding renormalized configuration $\sigma'_{i}$. In summary, based on the original system of lattice size $L_{0}$ we are able to extrapolate in parameter space observables of the renormalized system with $L^{(n)}_{L_{0}}$. Via reweighting, one obtains the values of renormalized observables that would emerge by applying the renormalization group transformation at a different point in parameter space: it is thus as if the transformation itself has been reweighted.

Before proceeding with the calculations of the critical fixed point and the critical exponents we will first probe what the permitted reweighting range is for the studied system. We express the expectation value of the action term that corresponds to the squared mass  $\langle S^{(1)} \rangle$  of the system in terms of each uniquely sampled value $S^{(1)}$ in the dataset, subject to a predefined bin size:
\begin{equation}
\langle S^{(1)} \rangle= \sum_{S^{(1)}} S^{(1)} w(S^{(1)}),
\end{equation}
where $w(S^{(1)})$ denotes the weight factor in the reweighting procedure which is defined as:
\begin{equation}
w(S^{(1)})= \frac{n(S^{(1)}) \exp[- \frac{1}{2}(\mu^{2}-\mu_{0}^{2})S^{(1)} ]}{\sum_{S^{(1)}}\exp[- \frac{1}{2}(\mu^{2}-\mu_{0}^{2})S^{(1)} ]},
\end{equation}
and $n(S^{(1)})$ are the histograms of $S^{(1)}$.

We are now able to calculate the weight factors $w(S^{(1)})$ defined above for different values of the reweighted squared mass $\mu^{2}$. This calculation is conducted for the larger lattice size $L_{0}=64$ since the reweighting range is anticipated to be smaller.  The results are depicted in Fig.~\ref{fig:fighist}, where the weight factor for the original coupling $\mu_{0}^{2}$, which is proportional to the actual histograms, is additionally included. We observe, based on the results in the figure, that weight factors can be reproduced for values in the range $\mu^{2} \in [-0.96,-0.94]$. However, inconsistencies emerge beyond that range, specifically for $\mu^{2}=-0.98$ and $\mu^{2}=-0.92$. Consequently, we consider a permitted reweighting range of $\mu^{2} \in [-0.96,-0.94]$, and we conduct calculations within this range.

\begin{figure}
\center
\includegraphics[width=8cm]{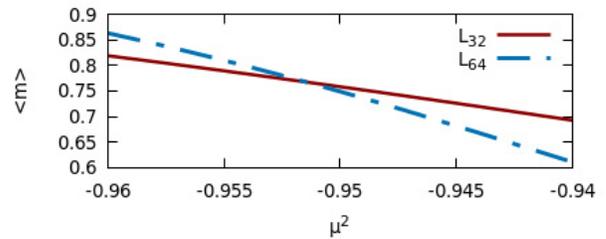}
\caption{\label{fig:figmagn} The magnetization $m$ versus the value of the squared mass $\mu^{2}$ for two renormalized systems of identical lattice size $L^{(4)}_{64}=L^{(3)}_{32}=4$. The width of each line indicates the statistical errors.}
\end{figure}

\section{Critical exponents and the critical fixed point}
\subsection{The correlation length exponent $\nu$}

Having obtained a permitted reweighting range we now depict the renormalized absolute value of the magnetization for two renormalized systems of identical lattice size  $L^{(4)}_{64}=L^{(3)}_{32}=4$  in Fig.~\ref{fig:figmagn}. The results are obtained via Eq.~(\ref{estimm}) based on reweighting from the original probability distributions. We observe that a critical fixed point is located since there exists an intersection between the two observables. 

In addition, we are able to associate to a certain value of the absolute magnetization $m$ a different squared mass for the two renormalized systems. This implies that we are able to use the data depicted in Fig.~\ref{fig:figmagn} to construct a mapping that relates the squared masses $\mu^{2}$ and $\mu'^{2}$ of the systems with lattice sizes  $L^{(3)}_{32}$ and $L^{(4)}_{64}$, respectively. This is achieved via the inverse mapping:
\begin{equation} \label{eq:mapping}
\mu'^{2}=O^{-1}(O'(\mu^{2})).
\end{equation}

\begin{figure}
\center
\includegraphics[width=8cm]{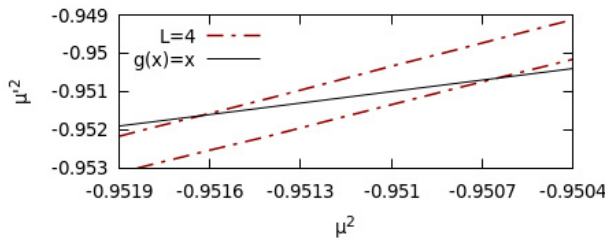}
\caption{\label{fig:figmap} The mappings which relate the squared masses $\mu^{2}$ and $\mu'^{2}$ of two renormalized systems of lattice sizes $L_{32}^{3}=4$ and $L_{64}^{4}=4$. The space bounded by the dashed lines depicts the statistical errors.}
\end{figure}

The results obtained from the inverse mapping are depicted in Fig.~\ref{fig:figmap}. The intersection with $g(x)=x$ provides a quantitative estimation of the critical fixed point $\mu_{c}^{2}=-0.95114(49)$, which agrees favorably with relevant literature~\cite{PhysRevD.79.056008}. The calculation considers fit and interpolation errors.

We remark that certain steps towards minimizing the systematic errors which emerge in computational renormalization group methods have been taken. Finite-size effects have been reduced by conducting the calculations on two systems of identical lattice size. Systematic errors pertinent to the choice of the transformation have been mitigated by the optimization process to determine the factor $f$. Errors included by the distance from the critical fixed point are alleviated due to the use of reweighting which enables an accurate determination of the fixed point by the intersection of observables. Another contribution of systematic errors emerges due to the choice of the operators used in the implementation of the renormalization group. We have only considered as an operator the magnetization $m$. To probe the magnitude of the latter systematic errors and to demonstrate that the results are consistent we now cross-verify the above calculation using as an additional operator the nearest-neighbor interaction $O_{ \phi_{i} \phi_{j}}$.

\begin{figure}
\center
\includegraphics[width=8cm]{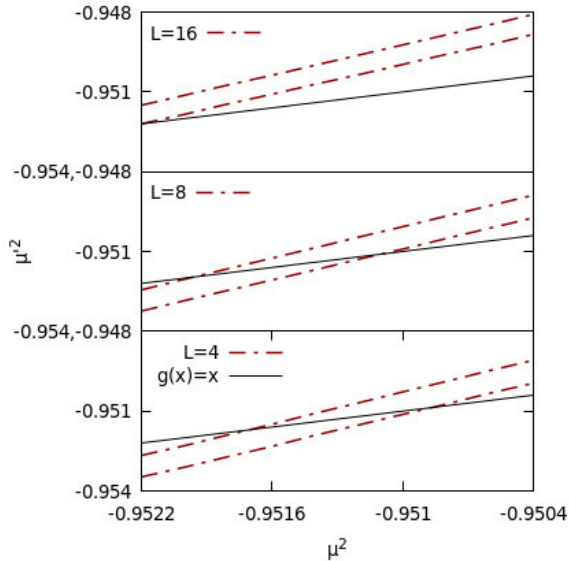}
\caption{\label{fig:fig3} The squared masses $\mu'^{2}$ and $\mu^{2}$ of two renormalized systems $L^{(n+1)}_{64}$ and $L^{(n)}_{32}$ for $n=1, 2, 3$ iterative applications of a renormalization group transformation (top to bottom, respectively). The intersection with $g(x)=x$ provides an estimation of the critical fixed point. The space bounded by the dashed lines indicates the statistical uncertainty.  }
\end{figure}

The inverse mappings for the two renormalized systems with $L^{(n+1)}_{64}=L^{(n)}_{32}$ for $n=1,2,3$, using the observable $O_{ \phi_{i} \phi_{j}}$, are depicted in Fig.~\ref{fig:fig3}.  Based on the lattice size $L^{(4)}_{64}=L^{(3)}_{32}=4$, we calculate the critical fixed point as $\mu_{c}^{2}=-0.95129(44)$, which agrees within statistical errors with the previous estimation. Consequently, the accurate calculation of the critical fixed point by two different observables indicates that the obtained result is consistent. 

 We now use the determined renormalized coupling parameters to extract the correlation length exponent $\nu$. The correlation lengths $\xi$ and $\xi'$ diverge in relation to the squared masses $\mu^{2}$ and $\mu'^{2}$ based on a critical exponent $\nu$ as:
\begin{equation}
\xi \sim |t|^{-\nu} ,
\end{equation}
\begin{equation}
 \xi' \sim |t'|^{-\nu}.
\end{equation}

We emphasize that even though the correlation lengths differ between the two renormalized systems, the correlation length exponent $\nu$ is identical since both systems are $\phi^{4}$ scalar field theories. By dividing the two expressions above, substituting the reduced coupling constants, linearizing the renormalization group transformation with a Taylor expansion to leading order, and taking the natural logarithm, we obtain:
\begin{equation}\label{eq:correxp}
\nu= \frac{\ln b}{\ln \frac{d\mu'^{2}}{d\mu^{2}} \Big|_{\mu^{2}_{c}}}.
\end{equation}

Based on the above expression, and the data depicted in Fig.~\ref{fig:figmap} and Fig.~\ref{fig:fig3}, we numerically calculate the correlation length exponent as $\nu=0.995(5)$ and $\nu=1.007(7)$, respectively. We remark that the phase transition of the two-dimensional $\phi^{4}$ scalar field theory is conjectured to be in the Ising universality class, hence the anticipated value of the critical exponent is $\nu=1$. 

\subsection{The magnetic field exponent $\theta$}

We now calculate the magnetic field exponent $\theta$ which governs the divergence of the correlation lengths $\xi$ and $\xi'$ in relation to the magnetic external field $h$ and $h'$. The analogous expressions are:
\begin{equation}
\xi \sim h^{-\theta} ,
\end{equation}
\begin{equation}
 \xi' \sim h'^{-\theta}.
\end{equation}

We define a modified lattice action $S^{(h)}$ for the $\phi^{4}$ theory which includes a coupled magnetic external field $h$:
\begin{equation}
S^{(h)}=S-h\sum_{i} \phi_{i}.
\end{equation}

While one would generally require Monte Carlo simulations to obtain configurations for a nonzero value of the magnetic external field $h$, we can instead implement reweighting to obtain the expectation values of observables that would correspond to the lattice action $S^{(h)}$. This can be achieved by using configurations sampled instead for the lattice action $S$. To introduce this reweighting method, and in line with Eq.~(\ref{estimo}), we define the expectation value $\langle O \rangle$ of an observable $O$ for the action $S^{(h)}$ as:
\begin{equation} \label{estimh}
\langle O \rangle=\frac{\sum_{i=1}^{N} O_{\sigma_{i}} \tilde{p}_{\sigma_i}^{-1}  \exp[-S_{\sigma_{i}}^{(h)} ]}{\sum_{i=1}^{N} \tilde{p}_{\sigma_i}^{-1}  \exp[ -S_{\sigma_{i}}^{(h)}]},
\end{equation}
and we again substitute $\tilde{p}$ as the probabilities of the original action $S$, to obtain:
\begin{equation} \label{estimh2}
\langle O \rangle=\frac{\sum_{i=1}^{N} O_{\sigma_{i}} \exp[ h \sum_{j} \phi_{j}^{(\sigma_{i})} ]}{\sum_{i=1}^{N} \exp[h \sum_{j} \phi_{j}^{(\sigma_{i})}]}.
\end{equation}

We observe that the above reweighting equation is agnostic to the lattice action~\cite{bachtis2020adding} and can, in principle, be used even when the form of the action is unknown. For completeness, we include the expression that enables the reweighting of a renormalized observable:

\begin{equation} \label{estimh23}
\langle O' \rangle=\frac{\sum_{i=1}^{N} O'_{\sigma'_{i}} \exp[ h \sum_{j} \phi_{j}^{(\sigma_{i})} ]}{\sum_{i=1}^{N} \exp[h \sum_{j} \phi_{j}^{(\sigma_{i})}]}.
\end{equation}

Based on the above expression, we observe that to reweight a renormalized observable $O'$ to a nonzero magnetic field $h$ we require only a calculation of $\sum_{j} \phi_{j}^{(\sigma_{i})}$ in the set of the original configurations $\sigma_{i}$.

Following the discussion pertinent to the construction of inverse mappings for the squared mass, we are now able to construct an inverse mapping which relates the original and the renormalized magnetic external fields $h$ and $h'$ as:
\begin{equation}
h'=O^{-1}(O'(h)).
\end{equation}

\begin{figure}
\center
\includegraphics[width=8.cm]{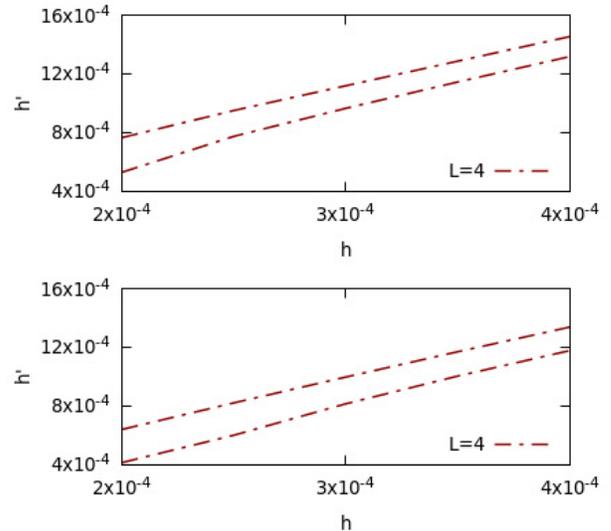}
\caption{\label{fig:figmaph} Magnetic fields $h'$ and $h$ for two renormalized systems of lattice size $L^{(4)}_{64}$ and $L^{(3)}_{32}$, respectively. The results are obtained from the construction of inverse mappings based on the magnetization (top) and the nearest neighbor interaction (bottom). The space bounded by the dashed lines depicts the statistical errors.}
\end{figure}

We are then able to calculate the magnetic field exponent $\theta$ as:
\begin{equation}\label{eq:fieldexp}
\theta= \frac{\ln b}{\ln \frac{d h'}{d h} \Big|_{h \rightarrow 0}}.
\end{equation}

The mappings between the two magnetic external fields, based on the observables of the magnetization and the nearest-neighbor interaction are depicted in Fig.~\ref{fig:figmaph}.  We remark that the configurations of the original actions were sampled in the vicinity of the phase transition so the original systems reside near the critical point $\mu_{c}^{2}$. The calculation of the magnetic field exponent $\theta$ is therefore conducted as $h\rightarrow 0^{+}$, so the systems remain at criticality.  We obtain the values $\theta=0.541(11)$ and $\theta=0.534(6)$ based on the magnetization and the nearest-neighbor interaction, respectively. The results agree favorably with the anticipated value of the two-dimensional Ising universality class $\theta=8/15$. Given the two exponents $\nu$ and $\theta$ the remaining exponents can be calculated via the use of scaling relations~\cite{newmanb99}. We emphasize that the above results were obtained with the use of histogram reweighting on original and renormalized observables without the need to simulate an action that includes a magnetic field $h$.

The results of this manuscript are obtained based on a given number of statistical samples, a determined factor $f$ in the use of a linear renormalization group transformation, the number of iterative renormalization group transformations, and the use of two operators to cross-verify the calculations. With this setting, the dominant source of uncertainty appears to be statistical. Nevertheless, this observation could change, either by considering a larger number of operators, of if the number of statistical samples is increased. The dominant source of uncertainty could then emerge as systematic. This systematic uncertainty could then be reduced based on a variety of approaches, for instance by further optimizing the renormalization group transformation.

To illustrate the computational benefits provided by histogram reweighting in the two-lattice matching Monte Carlo renormalization group calculations, consider the depicted range of $ \mu^{2} \in [-0.96,-0.94]$ in the results of Fig.~\ref{fig:figmagn}. These are obtained from two distinct systems with original lattice sizes $L_{0}=64$ and $L_{0}=32$. If one is interested in reproducing the results of the same figure with an accuracy of $10^{-5}$ using direct Monte Carlo simulations, one requires $4000$ Monte Carlo simulations. To each configuration from these $4000$ Monte Carlo simulations one would then need to apply a real-space transformation for $n$ iterations. In contrast, via the use of histogram reweighting and a step of $10^{-5}$ one is able to obtain the same results using only $2$ Monte Carlo simulations. To further observe the discussed benefits, one can additionally consider the results of Fig.~\ref{fig:figmaph}, which have been obtained via reweighting without the need to simulate an action with an explicit symmetry-breaking term.

\section{Conclusions}

We have shown that the incorporation of histogram reweighting within the two-lattice matching Monte Carlo renormalization group method provides computational benefits in the study of phase transitions. Specifically, instead of relying on a large number of direct Monte Carlo simulations  to construct an equivalence between observables of renormalized systems we can instead implement reweighting on exclusively two Monte Carlo simulations to accurately determine the renormalized coupling parameters in an extended region of parameter space. Furthermore, once the critical fixed point is discovered, an additional application of histogram reweighting with a reduced reweighting step can further refine the accuracy of the computational results.

The current work documents the application of histogram reweighting to determine the renormalized coupling parameters of a quantum field theory. Since the calculation of the critical exponents $\nu$ and $\theta$ which are associated with the relevant operators of the renormalization group is dependent on the accurate determination of the renormalized coupling parameters, the current work provides the first reweighting-enhanced calculation of these exponents for a quantum field theory. This is achieved by conducting the calculations on two systems of identical lattice size, thus reducing finite-size effects. We emphasize that, if required, the reweighting range can always be increased, in arbitrary dimensions, with the use of the multiple histogram method~\cite{PhysRevLett.63.1195}.

To conclude, in addition to computational benefits, the incorporation of histogram reweighting in two-lattice matching Monte Carlo renormalization group calculations is capable of introducing novel research directions. For instance, reweighting can be utilized to study complex-valued probability distributions~\cite{FODOR200287}, which cannot be directly simulated with importance sampling techniques. As a result, the possibility of extending Monte Carlo renormalization group methods to complex-valued probability distributions with the use of histogram reweighting is now an open research direction.

\section*{\label{sec:level5}Acknowledgements}
The author acknowledges support from the CFM-ENS Data Science Chair. Part of this research was conducted while the author was affiliated with Swansea University and the author acknowledges funding from the European Research Council (ERC) and the European Union's Horizon 2020 research and innovation programme under Grant Agreement No. 813942. 



\bibliography{ms}

\providecommand{\noopsort}[1]{}\providecommand{\singleletter}[1]{#1}%
\begin{thebibliography}{27}%
\makeatletter
\providecommand \@ifxundefined [1]{%
 \@ifx{#1\undefined}
}%
\providecommand \@ifnum [1]{%
 \ifnum #1\expandafter \@firstoftwo
 \else \expandafter \@secondoftwo
 \fi
}%
\providecommand \@ifx [1]{%
 \ifx #1\expandafter \@firstoftwo
 \else \expandafter \@secondoftwo
 \fi
}%
\providecommand \natexlab [1]{#1}%
\providecommand \enquote  [1]{``#1''}%
\providecommand \bibnamefont  [1]{#1}%
\providecommand \bibfnamefont [1]{#1}%
\providecommand \citenamefont [1]{#1}%
\providecommand \href@noop [0]{\@secondoftwo}%
\providecommand \href [0]{\begingroup \@sanitize@url \@href}%
\providecommand \@href[1]{\@@startlink{#1}\@@href}%
\providecommand \@@href[1]{\endgroup#1\@@endlink}%
\providecommand \@sanitize@url [0]{\catcode `\\12\catcode `\$12\catcode
  `\&12\catcode `\#12\catcode `\^12\catcode `\_12\catcode `\%12\relax}%
\providecommand \@@startlink[1]{}%
\providecommand \@@endlink[0]{}%
\providecommand \url  [0]{\begingroup\@sanitize@url \@url }%
\providecommand \@url [1]{\endgroup\@href {#1}{\urlprefix }}%
\providecommand \urlprefix  [0]{URL }%
\providecommand \Eprint [0]{\href }%
\providecommand \doibase [0]{https://doi.org/}%
\providecommand \selectlanguage [0]{\@gobble}%
\providecommand \bibinfo  [0]{\@secondoftwo}%
\providecommand \bibfield  [0]{\@secondoftwo}%
\providecommand \translation [1]{[#1]}%
\providecommand \BibitemOpen [0]{}%
\providecommand \bibitemStop [0]{}%
\providecommand \bibitemNoStop [0]{.\EOS\space}%
\providecommand \EOS [0]{\spacefactor3000\relax}%
\providecommand \BibitemShut  [1]{\csname bibitem#1\endcsname}%
\let\auto@bib@innerbib\@empty
\bibitem [{\citenamefont {Kadanoff}(1966)}]{PhysicsPhysiqueFizika.2.263}%
  \BibitemOpen
  \bibfield  {author} {\bibinfo {author} {\bibfnamefont {L.~P.}\ \bibnamefont
  {Kadanoff}},\ }\bibfield  {title} {\bibinfo {title} {Scaling laws for ising
  models near ${T}_{c}$},\ }\href
  {https://doi.org/10.1103/PhysicsPhysiqueFizika.2.263} {\bibfield  {journal}
  {\bibinfo  {journal} {Physics Physique Fizika}\ }\textbf {\bibinfo {volume}
  {2}},\ \bibinfo {pages} {263} (\bibinfo {year} {1966})}\BibitemShut {NoStop}%
\bibitem [{\citenamefont {Wilson}(1971)}]{PhysRevB.4.3174}%
  \BibitemOpen
  \bibfield  {author} {\bibinfo {author} {\bibfnamefont {K.~G.}\ \bibnamefont
  {Wilson}},\ }\bibfield  {title} {\bibinfo {title} {Renormalization group and
  critical phenomena. i. renormalization group and the kadanoff scaling
  picture},\ }\href {https://doi.org/10.1103/PhysRevB.4.3174} {\bibfield
  {journal} {\bibinfo  {journal} {Phys. Rev. B}\ }\textbf {\bibinfo {volume}
  {4}},\ \bibinfo {pages} {3174} (\bibinfo {year} {1971})}\BibitemShut
  {NoStop}%
\bibitem [{\citenamefont {Wilson}\ and\ \citenamefont
  {Fisher}(1972)}]{PhysRevLett.28.240}%
  \BibitemOpen
  \bibfield  {author} {\bibinfo {author} {\bibfnamefont {K.~G.}\ \bibnamefont
  {Wilson}}\ and\ \bibinfo {author} {\bibfnamefont {M.~E.}\ \bibnamefont
  {Fisher}},\ }\bibfield  {title} {\bibinfo {title} {Critical exponents in 3.99
  dimensions},\ }\href {https://doi.org/10.1103/PhysRevLett.28.240} {\bibfield
  {journal} {\bibinfo  {journal} {Phys. Rev. Lett.}\ }\textbf {\bibinfo
  {volume} {28}},\ \bibinfo {pages} {240} (\bibinfo {year} {1972})}\BibitemShut
  {NoStop}%
\bibitem [{\citenamefont {Wilson}\ and\ \citenamefont
  {Kogut}(1974)}]{WILSON197475}%
  \BibitemOpen
  \bibfield  {author} {\bibinfo {author} {\bibfnamefont {K.~G.}\ \bibnamefont
  {Wilson}}\ and\ \bibinfo {author} {\bibfnamefont {J.}~\bibnamefont {Kogut}},\
  }\bibfield  {title} {\bibinfo {title} {The renormalization group and the
  $\epsilon$ expansion},\ }\href
  {https://doi.org/https://doi.org/10.1016/0370-1573(74)90023-4} {\bibfield
  {journal} {\bibinfo  {journal} {Physics Reports}\ }\textbf {\bibinfo {volume}
  {12}},\ \bibinfo {pages} {75 } (\bibinfo {year} {1974})}\BibitemShut
  {NoStop}%
\bibitem [{\citenamefont {Wilson}(1975)}]{RevModPhys.47.773}%
  \BibitemOpen
  \bibfield  {author} {\bibinfo {author} {\bibfnamefont {K.~G.}\ \bibnamefont
  {Wilson}},\ }\bibfield  {title} {\bibinfo {title} {The renormalization group:
  Critical phenomena and the kondo problem},\ }\href
  {https://doi.org/10.1103/RevModPhys.47.773} {\bibfield  {journal} {\bibinfo
  {journal} {Rev. Mod. Phys.}\ }\textbf {\bibinfo {volume} {47}},\ \bibinfo
  {pages} {773} (\bibinfo {year} {1975})}\BibitemShut {NoStop}%
\bibitem [{\citenamefont {Newman}\ and\ \citenamefont
  {Barkema}(1999)}]{newmanb99}%
  \BibitemOpen
  \bibfield  {author} {\bibinfo {author} {\bibfnamefont {M.~E.~J.}\
  \bibnamefont {Newman}}\ and\ \bibinfo {author} {\bibfnamefont {G.~T.}\
  \bibnamefont {Barkema}},\ }\href@noop {} {\emph {\bibinfo {title} {Monte
  Carlo methods in statistical physics}}}\ (\bibinfo  {publisher} {Clarendon
  Press},\ \bibinfo {address} {Oxford},\ \bibinfo {year} {1999})\BibitemShut
  {NoStop}%
\bibitem [{\citenamefont {Ma}(1976)}]{PhysRevLett.37.461}%
  \BibitemOpen
  \bibfield  {author} {\bibinfo {author} {\bibfnamefont {S.-k.}\ \bibnamefont
  {Ma}},\ }\bibfield  {title} {\bibinfo {title} {Renormalization group by monte
  carlo methods},\ }\href {https://doi.org/10.1103/PhysRevLett.37.461}
  {\bibfield  {journal} {\bibinfo  {journal} {Phys. Rev. Lett.}\ }\textbf
  {\bibinfo {volume} {37}},\ \bibinfo {pages} {461} (\bibinfo {year}
  {1976})}\BibitemShut {NoStop}%
\bibitem [{\citenamefont {Swendsen}(1979)}]{PhysRevLett.42.859}%
  \BibitemOpen
  \bibfield  {author} {\bibinfo {author} {\bibfnamefont {R.~H.}\ \bibnamefont
  {Swendsen}},\ }\bibfield  {title} {\bibinfo {title} {Monte carlo
  renormalization group},\ }\href {https://doi.org/10.1103/PhysRevLett.42.859}
  {\bibfield  {journal} {\bibinfo  {journal} {Phys. Rev. Lett.}\ }\textbf
  {\bibinfo {volume} {42}},\ \bibinfo {pages} {859} (\bibinfo {year}
  {1979})}\BibitemShut {NoStop}%
\bibitem [{\citenamefont {Swendsen}(1982{\natexlab{a}})}]{Burkhardt1982}%
  \BibitemOpen
  \bibfield  {author} {\bibinfo {author} {\bibfnamefont {R.~H.}\ \bibnamefont
  {Swendsen}},\ }in\ \href {https://doi.org/10.1007/978-3-642-81825-7_1} {\emph
  {\bibinfo {booktitle} {Real-Space Renormalization}}},\ \bibinfo {editor}
  {edited by\ \bibinfo {editor} {\bibfnamefont {T.~W.}\ \bibnamefont
  {Burkhardt}}\ and\ \bibinfo {editor} {\bibfnamefont {J.~M.~J.}\ \bibnamefont
  {van Leeuwen}}}\ (\bibinfo  {publisher} {Springer Berlin Heidelberg},\
  \bibinfo {address} {Berlin, Heidelberg},\ \bibinfo {year} {1982})\BibitemShut
  {NoStop}%
\bibitem [{\citenamefont {Swendsen}(1982{\natexlab{b}})}]{swecargese}%
  \BibitemOpen
  \bibfield  {author} {\bibinfo {author} {\bibfnamefont {R.~H.}\ \bibnamefont
  {Swendsen}},\ }\href@noop {} {\emph {\bibinfo {title} {Phase transitions :
  Cargese 1980 / edited by Maurice Levy and Jean-Claude Le Guillou and Jean
  Zinn-Justin}}}\ (\bibinfo  {publisher} {Plenum Press : NATO Scientific
  Affairs Division New York},\ \bibinfo {year} {1982})\BibitemShut {NoStop}%
\bibitem [{\citenamefont {Wilson}(1980)}]{Wilsonbook}%
  \BibitemOpen
  \bibfield  {author} {\bibinfo {author} {\bibfnamefont {K.~G.}\ \bibnamefont
  {Wilson}},\ }in\ \href
  {https://doi.org/https://doi.org/10.1007/978-1-4684-7571-5} {\emph {\bibinfo
  {booktitle} {Recent Developments in Gauge Theories}}},\ \bibinfo {editor}
  {edited by\ \bibinfo {editor} {\bibfnamefont {G.}~\bibnamefont {Hooft}},
  \bibinfo {editor} {\bibfnamefont {C.}~\bibnamefont {Itzykson}}, \bibinfo
  {editor} {\bibfnamefont {A.}~\bibnamefont {Jaffe}}, \bibinfo {editor}
  {\bibfnamefont {H.}~\bibnamefont {Lehmann}}, \bibinfo {editor} {\bibfnamefont
  {P.~K.}\ \bibnamefont {Mitter}}, \bibinfo {editor} {\bibfnamefont {I.~M.}\
  \bibnamefont {Singer}},\ and\ \bibinfo {editor} {\bibfnamefont
  {R.}~\bibnamefont {Stora}}}\ (\bibinfo  {publisher} {Springer New York},\
  \bibinfo {address} {New York},\ \bibinfo {year} {1980})\BibitemShut {NoStop}%
\bibitem [{\citenamefont {Pawley}\ \emph {et~al.}(1984)\citenamefont {Pawley},
  \citenamefont {Swendsen}, \citenamefont {Wallace},\ and\ \citenamefont
  {Wilson}}]{PhysRevB.29.4030}%
  \BibitemOpen
  \bibfield  {author} {\bibinfo {author} {\bibfnamefont {G.~S.}\ \bibnamefont
  {Pawley}}, \bibinfo {author} {\bibfnamefont {R.~H.}\ \bibnamefont
  {Swendsen}}, \bibinfo {author} {\bibfnamefont {D.~J.}\ \bibnamefont
  {Wallace}},\ and\ \bibinfo {author} {\bibfnamefont {K.~G.}\ \bibnamefont
  {Wilson}},\ }\bibfield  {title} {\bibinfo {title} {{M}onte {C}arlo
  renormalization-group calculations of critical behavior in the simple-cubic
  {I}sing model},\ }\href {https://doi.org/10.1103/PhysRevB.29.4030} {\bibfield
   {journal} {\bibinfo  {journal} {Phys. Rev. B}\ }\textbf {\bibinfo {volume}
  {29}},\ \bibinfo {pages} {4030} (\bibinfo {year} {1984})}\BibitemShut
  {NoStop}%
\bibitem [{\citenamefont {Bowler}\ \emph {et~al.}(1986)\citenamefont {Bowler},
  \citenamefont {Hasenfratz}, \citenamefont {Hasenfratz}, \citenamefont
  {Heller}, \citenamefont {Karsch}, \citenamefont {Kenway}, \citenamefont
  {Pawley},\ and\ \citenamefont {Wallace}}]{BOWLER1986375}%
  \BibitemOpen
  \bibfield  {author} {\bibinfo {author} {\bibfnamefont {K.}~\bibnamefont
  {Bowler}}, \bibinfo {author} {\bibfnamefont {A.}~\bibnamefont {Hasenfratz}},
  \bibinfo {author} {\bibfnamefont {P.}~\bibnamefont {Hasenfratz}}, \bibinfo
  {author} {\bibfnamefont {U.}~\bibnamefont {Heller}}, \bibinfo {author}
  {\bibfnamefont {F.}~\bibnamefont {Karsch}}, \bibinfo {author} {\bibfnamefont
  {R.}~\bibnamefont {Kenway}}, \bibinfo {author} {\bibfnamefont
  {G.}~\bibnamefont {Pawley}},\ and\ \bibinfo {author} {\bibfnamefont
  {D.}~\bibnamefont {Wallace}},\ }\bibfield  {title} {\bibinfo {title} {The
  {SU(3)} $\beta$-function at large $\beta$},\ }\href
  {https://doi.org/https://doi.org/10.1016/0370-2693(86)90496-X} {\bibfield
  {journal} {\bibinfo  {journal} {Physics Letters B}\ }\textbf {\bibinfo
  {volume} {179}},\ \bibinfo {pages} {375} (\bibinfo {year}
  {1986})}\BibitemShut {NoStop}%
\bibitem [{\citenamefont {Hasenfratz}(2012)}]{PhysRevLett.108.061601}%
  \BibitemOpen
  \bibfield  {author} {\bibinfo {author} {\bibfnamefont {A.}~\bibnamefont
  {Hasenfratz}},\ }\bibfield  {title} {\bibinfo {title} {Infrared fixed point
  of the 12-fermion su(3) gauge model based on 2-lattice monte carlo
  renomalization-group matching},\ }\href
  {https://doi.org/10.1103/PhysRevLett.108.061601} {\bibfield  {journal}
  {\bibinfo  {journal} {Phys. Rev. Lett.}\ }\textbf {\bibinfo {volume} {108}},\
  \bibinfo {pages} {061601} (\bibinfo {year} {2012})}\BibitemShut {NoStop}%
\bibitem [{\citenamefont {Ferrenberg}\ and\ \citenamefont
  {Swendsen}(1988)}]{PhysRevLett.61.2635}%
  \BibitemOpen
  \bibfield  {author} {\bibinfo {author} {\bibfnamefont {A.~M.}\ \bibnamefont
  {Ferrenberg}}\ and\ \bibinfo {author} {\bibfnamefont {R.~H.}\ \bibnamefont
  {Swendsen}},\ }\bibfield  {title} {\bibinfo {title} {New monte carlo
  technique for studying phase transitions},\ }\href
  {https://doi.org/10.1103/PhysRevLett.61.2635} {\bibfield  {journal} {\bibinfo
   {journal} {Phys. Rev. Lett.}\ }\textbf {\bibinfo {volume} {61}},\ \bibinfo
  {pages} {2635} (\bibinfo {year} {1988})}\BibitemShut {NoStop}%
\bibitem [{\citenamefont {Bachtis}\ \emph
  {et~al.}(2021{\natexlab{a}})\citenamefont {Bachtis}, \citenamefont {Aarts},\
  and\ \citenamefont {Lucini}}]{PhysRevD.103.074510}%
  \BibitemOpen
  \bibfield  {author} {\bibinfo {author} {\bibfnamefont {D.}~\bibnamefont
  {Bachtis}}, \bibinfo {author} {\bibfnamefont {G.}~\bibnamefont {Aarts}},\
  and\ \bibinfo {author} {\bibfnamefont {B.}~\bibnamefont {Lucini}},\
  }\bibfield  {title} {\bibinfo {title} {Quantum field-theoretic machine
  learning},\ }\href {https://doi.org/10.1103/PhysRevD.103.074510} {\bibfield
  {journal} {\bibinfo  {journal} {Phys. Rev. D}\ }\textbf {\bibinfo {volume}
  {103}},\ \bibinfo {pages} {074510} (\bibinfo {year}
  {2021}{\natexlab{a}})}\BibitemShut {NoStop}%
\bibitem [{\citenamefont {Milchev}\ \emph {et~al.}(1986)\citenamefont
  {Milchev}, \citenamefont {Heermann},\ and\ \citenamefont
  {Binder}}]{Milchev1986}%
  \BibitemOpen
  \bibfield  {author} {\bibinfo {author} {\bibfnamefont {A.}~\bibnamefont
  {Milchev}}, \bibinfo {author} {\bibfnamefont {D.~W.}\ \bibnamefont
  {Heermann}},\ and\ \bibinfo {author} {\bibfnamefont {K.}~\bibnamefont
  {Binder}},\ }\bibfield  {title} {\bibinfo {title} {Finite-size scaling
  analysis of the $\phi^{4}$ field theory on the square lattice},\ }\href
  {https://doi.org/10.1007/BF01011906} {\bibfield  {journal} {\bibinfo
  {journal} {Journal of Statistical Physics}\ }\textbf {\bibinfo {volume}
  {44}},\ \bibinfo {pages} {749} (\bibinfo {year} {1986})}\BibitemShut
  {NoStop}%
\bibitem [{\citenamefont {Gonzalez-Arroyo}\ and\ \citenamefont
  {Okawa}(1987)}]{PhysRevD.35.672}%
  \BibitemOpen
  \bibfield  {author} {\bibinfo {author} {\bibfnamefont {A.}~\bibnamefont
  {Gonzalez-Arroyo}}\ and\ \bibinfo {author} {\bibfnamefont {M.}~\bibnamefont
  {Okawa}},\ }\bibfield  {title} {\bibinfo {title} {Renormalized coupling
  constants by monte carlo methods},\ }\href
  {https://doi.org/10.1103/PhysRevD.35.672} {\bibfield  {journal} {\bibinfo
  {journal} {Phys. Rev. D}\ }\textbf {\bibinfo {volume} {35}},\ \bibinfo
  {pages} {672} (\bibinfo {year} {1987})}\BibitemShut {NoStop}%
\bibitem [{\citenamefont {Swendsen}(1984)}]{PhysRevLett.52.2321}%
  \BibitemOpen
  \bibfield  {author} {\bibinfo {author} {\bibfnamefont {R.~H.}\ \bibnamefont
  {Swendsen}},\ }\bibfield  {title} {\bibinfo {title} {Optimization of
  real-space renormalization-group transformations},\ }\href
  {https://doi.org/10.1103/PhysRevLett.52.2321} {\bibfield  {journal} {\bibinfo
   {journal} {Phys. Rev. Lett.}\ }\textbf {\bibinfo {volume} {52}},\ \bibinfo
  {pages} {2321} (\bibinfo {year} {1984})}\BibitemShut {NoStop}%
\bibitem [{\citenamefont {Bachtis}\ \emph {et~al.}(2022)\citenamefont
  {Bachtis}, \citenamefont {Aarts}, \citenamefont {Di~Renzo},\ and\
  \citenamefont {Lucini}}]{PhysRevLett.128.081603}%
  \BibitemOpen
  \bibfield  {author} {\bibinfo {author} {\bibfnamefont {D.}~\bibnamefont
  {Bachtis}}, \bibinfo {author} {\bibfnamefont {G.}~\bibnamefont {Aarts}},
  \bibinfo {author} {\bibfnamefont {F.}~\bibnamefont {Di~Renzo}},\ and\
  \bibinfo {author} {\bibfnamefont {B.}~\bibnamefont {Lucini}},\ }\bibfield
  {title} {\bibinfo {title} {Inverse renormalization group in quantum field
  theory},\ }\href {https://doi.org/10.1103/PhysRevLett.128.081603} {\bibfield
  {journal} {\bibinfo  {journal} {Phys. Rev. Lett.}\ }\textbf {\bibinfo
  {volume} {128}},\ \bibinfo {pages} {081603} (\bibinfo {year}
  {2022})}\BibitemShut {NoStop}%
\bibitem [{\citenamefont {Brower}\ and\ \citenamefont
  {Tamayo}(1989)}]{PhysRevLett.62.1087}%
  \BibitemOpen
  \bibfield  {author} {\bibinfo {author} {\bibfnamefont {R.~C.}\ \bibnamefont
  {Brower}}\ and\ \bibinfo {author} {\bibfnamefont {P.}~\bibnamefont
  {Tamayo}},\ }\bibfield  {title} {\bibinfo {title} {Embedded dynamics for
  ${\mathrm{\ensuremath{\varphi}}}^{4}$ theory},\ }\href
  {https://doi.org/10.1103/PhysRevLett.62.1087} {\bibfield  {journal} {\bibinfo
   {journal} {Phys. Rev. Lett.}\ }\textbf {\bibinfo {volume} {62}},\ \bibinfo
  {pages} {1087} (\bibinfo {year} {1989})}\BibitemShut {NoStop}%
\bibitem [{\citenamefont {Schaich}\ and\ \citenamefont
  {Loinaz}(2009)}]{PhysRevD.79.056008}%
  \BibitemOpen
  \bibfield  {author} {\bibinfo {author} {\bibfnamefont {D.}~\bibnamefont
  {Schaich}}\ and\ \bibinfo {author} {\bibfnamefont {W.}~\bibnamefont
  {Loinaz}},\ }\bibfield  {title} {\bibinfo {title} {Improved lattice
  measurement of the critical coupling in ${\ensuremath{\phi}}_{2}^{4}$
  theory},\ }\href {https://doi.org/10.1103/PhysRevD.79.056008} {\bibfield
  {journal} {\bibinfo  {journal} {Phys. Rev. D}\ }\textbf {\bibinfo {volume}
  {79}},\ \bibinfo {pages} {056008} (\bibinfo {year} {2009})}\BibitemShut
  {NoStop}%
\bibitem [{\citenamefont {Loinaz}\ and\ \citenamefont
  {Willey}(1998)}]{PhysRevD.58.076003}%
  \BibitemOpen
  \bibfield  {author} {\bibinfo {author} {\bibfnamefont {W.}~\bibnamefont
  {Loinaz}}\ and\ \bibinfo {author} {\bibfnamefont {R.~S.}\ \bibnamefont
  {Willey}},\ }\bibfield  {title} {\bibinfo {title} {Monte carlo simulation
  calculation of the critical coupling constant for two-dimensional continuum
  ${\ensuremath{\varphi}}^{4}$ theory},\ }\href
  {https://doi.org/10.1103/PhysRevD.58.076003} {\bibfield  {journal} {\bibinfo
  {journal} {Phys. Rev. D}\ }\textbf {\bibinfo {volume} {58}},\ \bibinfo
  {pages} {076003} (\bibinfo {year} {1998})}\BibitemShut {NoStop}%
\bibitem [{\citenamefont {Wolff}(1989)}]{PhysRevLett.62.361}%
  \BibitemOpen
  \bibfield  {author} {\bibinfo {author} {\bibfnamefont {U.}~\bibnamefont
  {Wolff}},\ }\bibfield  {title} {\bibinfo {title} {Collective monte carlo
  updating for spin systems},\ }\href
  {https://doi.org/10.1103/PhysRevLett.62.361} {\bibfield  {journal} {\bibinfo
  {journal} {Phys. Rev. Lett.}\ }\textbf {\bibinfo {volume} {62}},\ \bibinfo
  {pages} {361} (\bibinfo {year} {1989})}\BibitemShut {NoStop}%
\bibitem [{\citenamefont {Bachtis}\ \emph
  {et~al.}(2021{\natexlab{b}})\citenamefont {Bachtis}, \citenamefont {Aarts},\
  and\ \citenamefont {Lucini}}]{bachtis2020adding}%
  \BibitemOpen
  \bibfield  {author} {\bibinfo {author} {\bibfnamefont {D.}~\bibnamefont
  {Bachtis}}, \bibinfo {author} {\bibfnamefont {G.}~\bibnamefont {Aarts}},\
  and\ \bibinfo {author} {\bibfnamefont {B.}~\bibnamefont {Lucini}},\
  }\bibfield  {title} {\bibinfo {title} {Adding machine learning within
  hamiltonians: Renormalization group transformations, symmetry breaking and
  restoration},\ }\href {https://doi.org/10.1103/PhysRevResearch.3.013134}
  {\bibfield  {journal} {\bibinfo  {journal} {Phys. Rev. Research}\ }\textbf
  {\bibinfo {volume} {3}},\ \bibinfo {pages} {013134} (\bibinfo {year}
  {2021}{\natexlab{b}})}\BibitemShut {NoStop}%
\bibitem [{\citenamefont {Ferrenberg}\ and\ \citenamefont
  {Swendsen}(1989)}]{PhysRevLett.63.1195}%
  \BibitemOpen
  \bibfield  {author} {\bibinfo {author} {\bibfnamefont {A.~M.}\ \bibnamefont
  {Ferrenberg}}\ and\ \bibinfo {author} {\bibfnamefont {R.~H.}\ \bibnamefont
  {Swendsen}},\ }\bibfield  {title} {\bibinfo {title} {Optimized monte carlo
  data analysis},\ }\href {https://doi.org/10.1103/PhysRevLett.63.1195}
  {\bibfield  {journal} {\bibinfo  {journal} {Phys. Rev. Lett.}\ }\textbf
  {\bibinfo {volume} {63}},\ \bibinfo {pages} {1195} (\bibinfo {year}
  {1989})}\BibitemShut {NoStop}%
\bibitem [{\citenamefont {Fodor}\ and\ \citenamefont
  {Katz}(2002)}]{FODOR200287}%
  \BibitemOpen
  \bibfield  {author} {\bibinfo {author} {\bibfnamefont {Z.}~\bibnamefont
  {Fodor}}\ and\ \bibinfo {author} {\bibfnamefont {S.}~\bibnamefont {Katz}},\
  }\bibfield  {title} {\bibinfo {title} {A new method to study lattice {QCD} at
  finite temperature and chemical potential},\ }\href
  {https://doi.org/https://doi.org/10.1016/S0370-2693(02)01583-6} {\bibfield
  {journal} {\bibinfo  {journal} {Physics Letters B}\ }\textbf {\bibinfo
  {volume} {534}},\ \bibinfo {pages} {87} (\bibinfo {year} {2002})}\BibitemShut
  {NoStop}%
\end{thebibliography}%
\end{document}